\documentclass[12pt]{iopart}
\usepackage{graphics,bm}
\usepackage{amssymb,ulem}
\usepackage{epsfig}
\usepackage{epsf}
\usepackage[usenames]{color}

\begin{document}

\title[Spin Drag in Ultracold Fermi Mixtures with Repulsive Interactions]{Spin Drag in Ultracold Fermi Mixtures with Repulsive Interactions}

\author{R.A. Duine}
\address{Institute for Theoretical Physics, Utrecht
University, Leuvenlaan 4, 3584 CE Utrecht, The Netherlands}

\author{Marco Polini}
\address{NEST, Istituto Nanoscienze-CNR and Scuola Normale Superiore,
I-56126 Pisa, Italy}

\author{Arnaud Raoux}
\address{Formation Interuniversitaire de Physique,
D\'epartement de Physique de l'\'Ecole Normale Sup\'erieure, 24 rue Lhomond, 75231 Paris Cedex 05, France}

\author{H.T.C. Stoof}

\address{Institute for Theoretical Physics, Utrecht
University, Leuvenlaan 4, 3584 CE Utrecht, The Netherlands}

\author{G. Vignale}
\address{Department of Physics and Astronomy, University of
Missouri, Columbia, Missouri 65211, USA}

\begin{abstract}
We calculate the spin-drag relaxation rate for a two-component ultracold atomic Fermi gas with positive scattering length between the two spin components. In one dimension we find that it vanishes linearly with temperature. In three dimensions the spin-drag relaxation rate vanishes quadratically with temperature for sufficiently weak interactions. This quadratic temperature dependence is present, up to logarithmic corrections, in the two-dimensional case as well. For stronger interaction the system exhibits a Stoner ferromagnetic phase transition in two and three dimensions. We show that the spin-drag relaxation rate is enhanced by spin fluctuations as the temperature approaches the critical temperature of this transition from above.
\end{abstract}

%Uncomment for PACS numbers title message
%\pacs{00.00, 20.00, 42.10}
% Keywords required only for MST, PB, PMB, PM, JOA, JOB?
%\vspace{2pc}
%\noindent{\it Keywords}: Article preparation, IOP journals
% Uncomment for Submitted to journal title message
%\submitto{\JPA}
% Comment out if separate title page not required
\maketitle

% definitions
\def\bx{{\bm x}}
\def\bk{{\bm k}}
\def\bK{{\bm K}}
\def\bq{{\bm q}}
\def\br{{\bm r}}
\def\bp{{\bm p}}
\def\half{\frac{1}{2}}
\def\args{(\bm, t)}

\section{Introduction}
Interest in electronic transport ranges
from everyday applications to fundamental physics. One of the most
interesting phenomena that spans this entire range, is the
influence of a thermodynamic phase transition on the electrical
conductivity. The most direct example is the phase transition from
a normal conductor to a superconductor characterized by a
vanishing resistance. The applications of this phenomenon are
ubiquitous and the basic physics that underlies the transition in
superconductors, the Bose-Einstein condensation of fermionic
pairs, has emerged in research fields from astroparticle
physics~\cite{son1999} to cold-atom
systems~\cite{stoof1996,regal2004}.

A system in between the latter two temperature extremes, in which
analogies of superconductivity have been predicted, is that of a
two-dimensional electron-hole
bilayer~\cite{shevchenko1976,lozovik1976}. In this case the pairs
that condense are excitons formed by electrons from one layer with
holes in the other. The relevant transport probe is in this case
the Coulomb drag measurement~\cite{coulombdrag}: a current $I$ is
driven through one layer, known as the ``active" layer, causing a
voltage drop $V_{\rm D}$ in the other. As the layers are separated
by an essentially impenetrable tunnel barrier, the voltage drop is
predominantly caused by Coulomb scattering, and the drag
resistivity $\rho_{\rm D} = V_{\rm D} / I$ has the, up to logarithmic corrections, characteristic
quadratic Fermi-liquid-like low-temperature dependence $\rho_{\rm
D} \propto T^2$. When the excitons undergo Bose-Einstein
condensation, however, the drag resistivity is predicted to jump
from the relatively small value proportional to $T^2$ to a value
equal to the ordinary resistivity of the active
layer~\cite{vignale1996}. Although conclusive evidence of exciton
condensation is still lacking, two experimental
groups~\cite{croxall2008,seamons2009} have recently reported the
observation of an upturn in the drag resistivity as the
temperature is lowered. This upturn is interpreted as being due to
strong pairing fluctuations that precede exciton
condensation~\cite{hu2000} and thus serves as a precursor signal
for the transition, similar to the enhancement of the conductivity
in superconductors due to superconducting fluctuations above but
close to the critical temperature~\cite{larkin2002}.

A closely related situation arises when the two layers of a two-dimensional electron-electron bilayer placed in a strong perpendicular magnetic field are close enough to
allow the establishment of interlayer coherence~\cite{eisenstein_nature_2004}. In this case, the
two layers in the system can be labelled ``up" and ``down" along a
``z"-axis, so that the which-layer degree of freedom becomes a
spin one-half pseudospin. Interlayer coherence in this language
corresponds to pseudospin ferromagnetism with an easy x-y plane,
since this orientation of the pseudospin describes a particle that
is neither in the left nor in the right layer, but in a coherent
superposition of the two. Furthermore, Coulomb drag becomes
pseudospin drag, the mutual friction between two pseudospin states
due to Coulomb scattering. This analogy prompted studies of spin
drag, the frictional drag between electrons with opposite spin
projection, in a single semiconductor~\cite{scd_giovanni}. While
the realization of separate electric contacts to the two spin
states remains an experimentally challenging problem, the spin
drag is observed indirectly, by measuring different diffusion
constants for charge and spin~\cite{weber_nature_2005}.

Because of the presence of other relaxation mechanisms, spin-drag
effects are usually not very large in semiconductors, and are even
smaller in metals. This is completely different in cold atomic
gases where scattering between different hyperfine spins is the
only mechanism to relax spin currents, and was considered both for
fermionic atoms~\cite{polini_prl_2007,bruun_prl_2008}, and for
bosonic ones~\cite{duine_prl_2009}. In this paper we consider spin drag in a two-component Fermi gas, in one, two, and three dimensions. We point out that a particularly interesting situation
occurs when spin drag is considered in a two-component Fermi gas
that is close to a ferromagnetic
instability~\cite{houbiers1997,salasnich2000,amoruso2000,sogo2002,duine2005}, as can occur for sufficiently strong and repulsive interactions in two and three dimensions.
We show that the spin drag is strongly
enhanced as the ferromagnetic state is approached from the normal
side~\cite{duine2010}, as expected from the analogy between electron-hole bilayers and pseudospin
ferromagnets. In one dimension, however, where the ferromagnetic phase transition is absent, the effects of spin drag vanish linearly with temperature.
One of our
motivations for considering this effect is the recent observation
of ferromagnetic correlations in a two-component Fermi gas with
strong repulsive interactions~\cite{jo2009}. The fact that
spin-polarized domains were not directly observed adds to the
theoretical interest~\cite{conduit2009} in this experiment.
Because atoms are neutral, the relevant experimental
quantity is the spin-drag relaxation rate, which for instance
determines the damping rate of the spin-dipole mode in trapped
cold-atom systems~\cite{bruun_prl_2008} and is thus accessible experimentally. Interestingly, an electronic analog of
the spin-dipole mode also exists~\cite{damico_prb_2006}.

The remainder of this paper is organized as follows. We first derive an expression for the damping of the spin dipole mode from the Boltzmann equation. As mentioned before, this damping is determined by the spin-drag relaxation rate, which is subsequently evaluated in one, two, and three dimensions. We end with our conclusions and a short discussion.

\section{Spin dipole mode and spin-drag relaxation rate} \label{sec:spindipolemode}
We consider a mixture of fermionic atoms of mass $m$ in $d$ dimensions,
with two hyperfine states denoted by $|\uparrow\rangle$ and
$|\downarrow\rangle$. The grand-canonical Hamiltonian with external trapping potential $V(\bx)$
and chemical potential $\mu$ is given by
\begin{eqnarray}
{\hat H} &=& \int\!d^d\bx \sum_{\sigma\in\{\uparrow,\downarrow\}}
{\hat \psi}^\dagger_\sigma (\bx)\left(- \frac{\hbar^2 {\nabla^2_\bx}}{2m} +V (\bx) -\mu \right) {\hat \psi}_\sigma (\bx) \nonumber \\
&+& U \int\!d^d\bx~{\hat \psi}^\dagger_\uparrow (\bx) {\hat
\psi}^\dagger_\downarrow (\bx) {\hat \psi}_\downarrow (\bx) {\hat
\psi}_\uparrow(\bx)~,
\end{eqnarray}
in terms of fermionic creation and annihilation operators ${\hat
\psi}_\alpha^\dagger (\bx)$ and ${\hat \psi}_\alpha (\bx)$,
respectively. At low temperatures $s$-wave scattering, described
by a pseudopotential $V(\bx-\bx') = U \delta (\bx-\bx')$, dominates, and we have therefore
omitted other interaction terms from this Hamiltonian. We here consider only the balanced case in which there is an equal number of atoms $N$ in each hyperfine state.

Following the discussion in Ref.~\cite{vandriel2010} we now derive an expression for the damping of the spin-dipole mode from the Boltzmann equation for the distribution function $f_\sigma (\bx,\bk,t)$ for the atoms in spin state $|\sigma\rangle$, given by
\begin{equation}
\frac{\partial f_\uparrow}{\partial t}  -
\frac{1}{\hbar} \bm{\nabla} V \cdot \frac{\partial f_\uparrow}{\partial
\bm{k}} + \frac{\hbar \bm{k}}{m}\cdot \frac{\partial
f_\uparrow}{\partial \bm{x}} = \Gamma_{\rm coll}[f_\uparrow,f_\downarrow]~,
\end{equation}
where we take the trapping potential to be harmonic $V (\bx)=m \sum_{j=1}^d \omega_j^2 x_j^2/2$. The equation for $f_\downarrow$ is  found by replacing $f_\uparrow \leftrightarrow f_\downarrow$ in the above. Below we give an explicit expression for the collision integral $ \Gamma_{\rm coll}[f_\uparrow,f_\downarrow]$.

We solve this inhomogeneous Boltzmann equation by making the {\it
ansatz} $ f_\uparrow({\bm x}, {\bm k}, t) = N_{\rm F} (\epsilon_{{\bm k} - m
{\bm v}_\uparrow(t)/\hbar} + V({\bm x}-{\bm x}_\uparrow (t)))$, with a similar
expression for $f_\downarrow ({\bm x},{\bm k},t)$. Here, $\epsilon_\bk=\hbar^2 \bk^2/2m$ is the single-particle dispersion and $N_{\rm F}(\epsilon) = [e^{\beta (\epsilon-\mu)}+1]^{-1}$ is the Fermi-Dirac distribution function with $\beta = (k_{\rm B}T)^{-1}$ the inverse temperature. This {\it ansatz} is
parameterized by the center-of-mass velocity ${\bm v}_{\sigma} (t)$
and position ${\bm x}_{\sigma} (t)$ of the atomic cloud of atoms in spin state $|\sigma\rangle$. From
this, we get the equations of motion
\begin{eqnarray}\label{eq:EOM2}
N m \frac{d{\bm v}_\uparrow}{dt} =  - N
\frac{dV\!\left( {\bm x}_\uparrow\right)}{d{\bm x}_\uparrow} + {\bm \Gamma}({\bm v}_\uparrow
-
{\bm v}_\downarrow, {\bm x}_\downarrow - {\bm x}_\uparrow)~;\\
\nonumber N m \frac{d{\bm v}_\downarrow}{dt} = - N
\frac{dV\!\left( {\bm x}_\downarrow\right)}{d{\bm x}_\downarrow} - {\bm \Gamma}({\bm v}_\downarrow
- {\bm v}_\uparrow, {\bm x}_\downarrow - {\bm x}_\uparrow)~,
\end{eqnarray}
with the function ${\bm \Gamma}({\bm v}_\downarrow
- {\bm v}_\uparrow, {\bm x}_\downarrow - {\bm x}_\uparrow)$ given by
\begin{eqnarray}
&& {\bm \Gamma}({\bm v}_\downarrow
- {\bm v}_\uparrow, {\bm x}_\downarrow - {\bm x}_\uparrow) = \int d^d {\bm x} \int \frac{d^d \bm{k}}{(2\pi)^d} \hbar\bm{k} \nonumber \\
&&\times \Gamma_{\rm coll} \left[N_{\rm F} (\epsilon_{{\bm k} - m
{\bm v}_\downarrow(t)/\hbar} + V({\bm x}\!-\!{\bm x}_\downarrow (t))), N_{\rm F} (\epsilon_{{\bm k} - m {\bm v}_\uparrow(t)/\hbar} + V({\bm
x}\!-\!{\bm x}_\uparrow (t)))\right]~,
\nonumber \\
\end{eqnarray}
where
\begin{eqnarray*}
&& \hspace*{-0.4cm} \nonumber \Gamma_{\rm coll}[f_\downarrow,f_\uparrow] =
\frac{(2\pi)^{d+1}}{\hbar} U^2 \int \frac{d^d {\bm
k}_2}{(2\pi)^d} \int \frac{d^d {\bm
k}_3}{(2\pi)^d} \int \frac{d^d {\bm k}_4}{(2\pi)^d} \\
&&\times\delta^d({\bm k} + {\bm k}_2 - {\bm k}_3 - {\bm k}_4)
\delta(\epsilon_{\bm k} + \epsilon_{{\bm k}_2} - \epsilon_{{\bm
k}_3} - \epsilon_{{\bm k}_4}) \\ &&\times \{[1 - f_\uparrow({\bm x}, {\bm
k}, t)][1 - f_\downarrow({\bm x},{\bm k}_2, t)]f_\uparrow({\bm x},{\bm k}_3,
t)f_\downarrow({\bm x},{\bm k}_4, t)  \\ &&- f_\uparrow({\bm x},{\bm k},
t)f_\downarrow({\bm x},{\bm k}_2, t)[1 - f_\uparrow({\bm x}, {\bm k}_3, t)][1 -
f_\downarrow({\bm x},{\bm k}_4, t)]\}~.
\end{eqnarray*}
We linearize the above equations using that ${\bm
\Gamma}({\bm v}, {\bm x}) \simeq \Gamma' {\bm v}$ due to the
isotropy of the collision integral.
The linearized equations then yield a collective-mode spectrum with $2d$ modes, corresponding to two types of oscillation in the $d$-dimensional trap. One set of modes is undamped and has frequencies $\omega_j$, $j\in \{1,\ldots,d\}$, and corresponds to an in-phase oscillation of the two clouds in the harmonic trap.
The other mode corresponds to the out-of-phase spin dipole oscillation of the two spin states. This mode is damped as a result of the friction, {\it i.e.}, the spin drag between the two spin states during the oscillation. This friction is due to collisions between particles of opposite spin and results in
transfer of momentum between the two clouds, leading to spin drag
and damping of these modes. These modes have the frequencies
\begin{equation}
\omega^{\rm dip}_j = - i \gamma + \sqrt{\omega_j^2 -\gamma^2
}~.
\end{equation}
The imaginary part of the above frequencies gives the damping rate
of the modes, and is given  by $\gamma \equiv (2\tau_{\rm sd})^{-1} = \Gamma'/Nm$ with $\tau_{\rm sd}$ the
spin-drag relaxation time.

We proceed by giving an expression for $\tau_{\rm sd}$ for a
homogeneous system with density $n$ per spin state for which we, in first
approximation, have taken the central density in the trap to make
connection with the inhomogeneous case. In
terms of the noninteracting (Lindhard) response
function at nonzero temperature
\begin{equation}
\label{eq:lindhard} \chi_0(q,\omega) =2 \int \frac{d^d\bk}{(2\pi)^d}
\frac{N_{\rm F}(\epsilon_{\bq+\bk}) - N_{\rm F} (\epsilon_{\bk})} {\epsilon_{\bq+\bk} - \epsilon_{\bk}
-\hbar\omega -i0}~,
\end{equation}
the expression for $\Gamma'$ can be worked out to yield
\begin{equation}\label{eq:resulttausd}
\frac{1}{\tau_{\rm sd}(T)} = \frac{\hbar^2}{4 m n k_{\rm B} T}
\int \frac{d^d\bq}{(2\pi)^d} \frac{q^2}{d}
\int_{-\infty}^{+\infty} \frac{d\omega}{\pi}
U^2\frac{[\Im
m~\chi_0(q,\omega)]^2} {\sinh^2[\hbar \omega/(2 k_{\rm B} T)]}.
\end{equation}
In the next section we present results obtained by evaluating this expression. We end this section by noting that the above expression for
the spin-drag relaxation rate is similar to the expression for the drag resistivity in electron-hole bilayers \cite{coulombdrag}. This also implies that the
spin-drag relaxation rate defines a dissipative transport coefficient for a cold-atom system and thus represents a natural starting point for studying transport phenomena in these systems.

\section{Results for the spin-drag relaxation rate} \label{sec:resultsspindragrelaxationrate}

In this section we present results for the spin-drag relaxation rate $1/\tau_{\rm sd}$. These results are, among other parameters, characterized by the Fermi wave number $k_{\rm F} = [d \Gamma(d/2) n/4\pi^{d/2}]^{1/d}$, where $\Gamma(x)$ is the Euler Gamma function. We also introduce the Fermi energy $\varepsilon_{\rm F} = k_{\rm B}T_{\rm F} = \hbar^2
k_{\rm F}^2/2m$. The results for the one-dimensional (three-dimensional) case are also discussed in Ref.~\cite{polini_prl_2007} (Ref.~\cite{duine2010}).

\begin{figure}
\begin{center}
\includegraphics[width=0.6\linewidth]{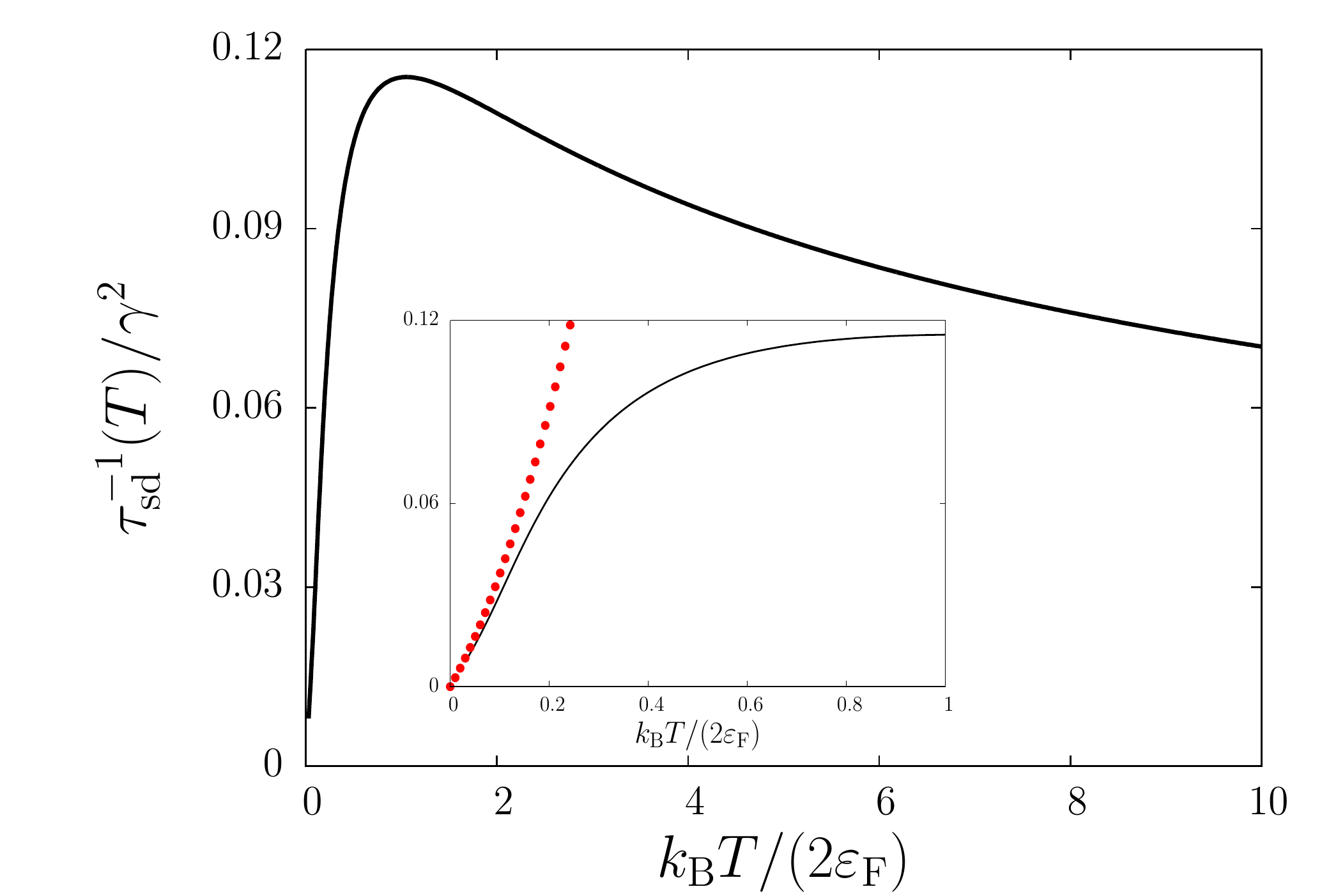}
\caption{ Spin-drag relaxation rate $\tau^{-1}_{\rm sd}$ (in units of $\varepsilon_{\rm F}/\hbar$) as a function of the temperature for a one-dimensional Fermi gas. The inset shows a zoom of the low-temperature region $0\leq k_{\rm B}T/2\varepsilon_{\rm F} \leq 1$, with the filled circles representing the analytical result in Eq.~(\ref{eq:low_t}). \label{fig:fig1}}
\end{center}
\end{figure}
\subsection{One dimension}

A one-dimensional trapped gas can be experimentally realized by tightly confining two directions in the harmonic trap. We therefore take $\omega_1=\omega_2 \equiv \omega_\perp$ to be much larger than $\omega_3$. In the limit $a \ll a_\perp$, where $a_\perp=\sqrt{\hbar^2/m\omega_\perp}$ and $a$ is the three-dimensional $s$-wave scattering length, one has for the effective one-dimensional coupling constant that $U_{\rm 1D}=2\hbar^2 a/ma^2_\perp$~\cite{olshanii_1998}. It is also customary to introduce the dimensionless Yang parameter $\gamma=m U_{\rm 1D}/\hbar^2 n$.

In Fig.~\ref{fig:fig1} we show the results that follow from Eq.~(\ref{eq:resulttausd}) by taking $d=1$ and $U=U_{1D}$. From this plot it is seen that the spin-drag relaxation rate vanishes linearly with temperature. It can be shown~\cite{polini_prl_2007} from Eq.~(\ref{eq:resulttausd}) that
\begin{equation}\label{eq:low_t}
\frac{1}{\tau_{\rm sd}(T)}\stackrel {T\to 0}{\rightarrow}\left[\frac{8}{9\pi}\gamma^2 \frac{k_{\rm B} T}{2\varepsilon_{\rm F}}
+\frac{8}{3\pi}\gamma^2\left(\frac{k_{\rm B} T}{2\varepsilon_{\rm F}}\right)^2\right]\frac{\varepsilon_{\rm F}}{\hbar}~,                                      \end{equation}
in agreement with the numerical results for $T/T_{\rm F} \ll 1$ (see the inset of Fig.~\ref{fig:fig1}). This leading order behavior is in agreement with results from bosonization~\cite{bosonization}. The linear-in-$T$ term in square brackets in Eq.~(\ref{eq:low_t}) originates from contributions to the spin-drag relaxation rate that are controlled by momenta $q$ of the order of $2 k_{\rm F}$, while the quadratic-in-$T$ term comes from momenta $q$ near $0$. Calculations beyond second-order perturbation theory have been carried out by Pustilnik {\it et al.}~\cite{pustilnik2003} in the context of Coulomb drag between quantum wires: these authors have considered only contributions to the drag transresistance $\rho_{\rm D}$ coming from momenta $q$ near $0$ and found $\rho_{D} \propto T^2$ for $T \to 0$. Neglecting the $2 k_{\rm F}$ contributions to $\rho_{\rm D}$ is fully justified in their case since the inter-wire Coulomb interaction at wave vectors of the order of $2k_{\rm F}$ is suppressed by the exponential factor $\exp(-2 k_{\rm F} \ell)$, where $\ell$ is the inter-wire distance.

\begin{figure}
\begin{center}
\includegraphics[width=0.6\linewidth]{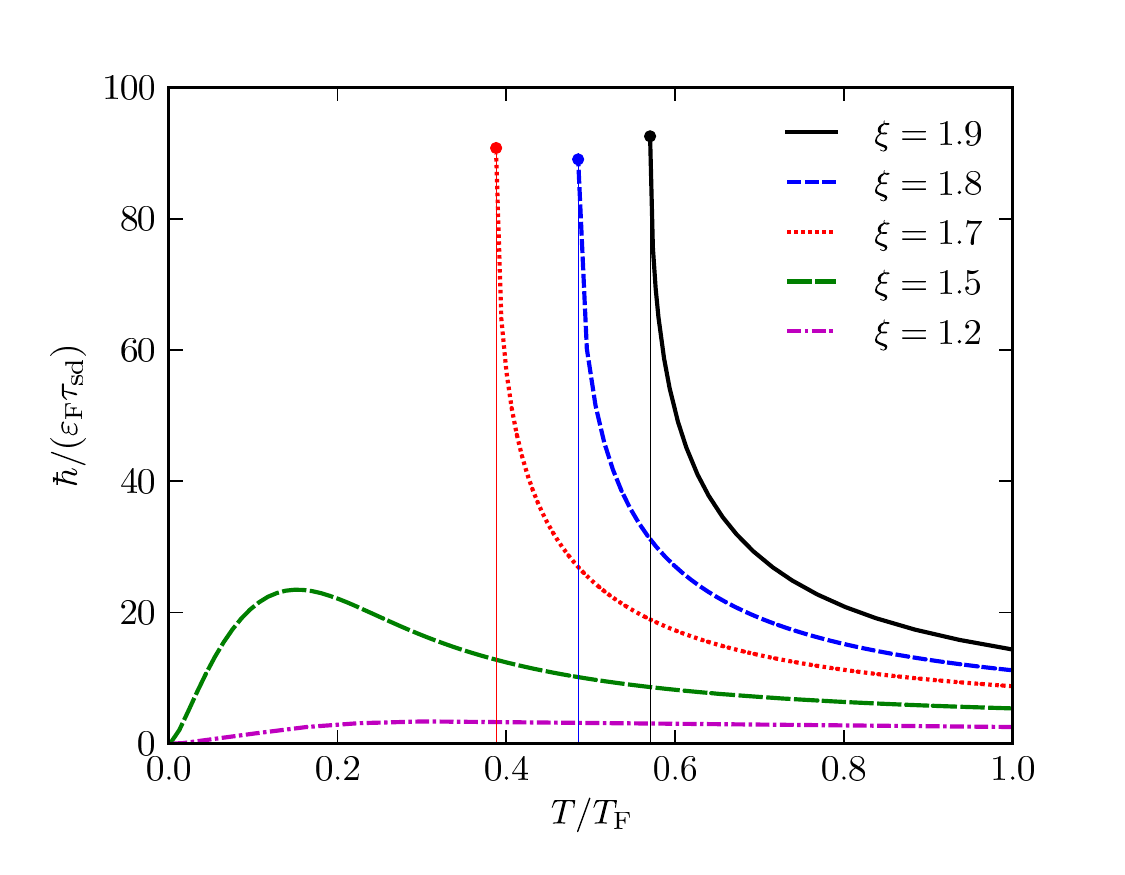}
\caption{Spin-drag relaxation rate in two dimensions as a function of temperature, and for various strengths of the interactions determined by $\xi \equiv \pi a/a_z$. The vertical lines indicate the critical temperature for the ferromagnetic phase transition. \label{fig:fig2}}
\end{center}
\end{figure}

\subsection{Two and three dimensions}
In two and three dimensions the spin one-half Fermi gas is predicted to undergo a ferromagnetic phase transition \cite{houbiers1997,salasnich2000,amoruso2000,sogo2002,duine2005}, which motivated the experiments by Jo {\it et al.} \cite{jo2009}. Assuming a second-order phase transition (note, however, that the transition is predicted to become first order at very low temperatures \cite{duine2005}), the transition is signalled by a diverging spin susceptibility $\chi_{S_zS_z} (q,\omega)$ at zero wave vector ($q=0$) and frequency ($\omega=0$). Within Stoner mean-field theory this spin susceptibility is calculated by summing all random-phase approximation (RPA) bubble diagrams, which yields $\chi_{S_zS_z} (q,\omega) = \chi_0(q,\omega)/[1+
U\chi_0(q,\omega)/2]$. Hence, the critical temperature, both in two and three dimensions, is determined by the condition $1+U \chi_0(0,0)/2=0$. This
equation gives, together with the equation $n = 2 \int
d^d\bq~N_{\rm F} (\epsilon_{\bq})/(2\pi)^d$ for the total density that determines the
chemical potential, the critical temperature $T_{\rm c}$ as a
function of $U$. In two dimensions this yields
\begin{equation}
  \frac{T_{\rm F}}{T_{\rm c}} + \log \left(1-e^{-\frac{T_{\rm F}}{T_{\rm c}}} \right) =
  - \log \left[ \frac{U \nu (\epsilon_{\rm F})}{2} -1 \right]~,
\end{equation}
with $\nu (\epsilon_{\rm F})$ the total density of states at the Fermi level. Note that only when $U \nu (\epsilon_F)/2>1$, the Stoner criterion, there exists a ferromagnetic phase transition.

Experimentally, the two-dimensional situation can be achieved by tightly confining the system in one direction by making one (say $\omega_z$) of the three trapping frequencies much larger than the other two. The effective two-dimensional interaction is then determined by $U_{2D} = 4 \pi a \hbar^2/m a_z$, where $a_z=\sqrt{\hbar/m\omega_z}$.

To account for the effect of ferromagnetic fluctuations, we evaluate the effective scattering amplitude between atoms by summing all RPA bubble diagrams. This gives
\begin{eqnarray}\label{eq:scattering_amplitude_transparent}
A_{\uparrow\downarrow}(q,\omega) = U_{2D} +
\frac{U_{2D}^2}{4}~\chi_{nn}(q,\omega) - \frac{3
U_{2D}^2}{4}~\chi_{S_zS_z}(q,\omega)~,
\end{eqnarray}
with $\chi_{nn}(q,\omega) =\chi_0(q,\omega)/[1-
U\chi_0(q,\omega)/2]$ the RPA density response function. In what follows we numerically evaluate the result in Eq.~(\ref{eq:resulttausd}) with the above effective interaction, {\it i.e.}, after making the replacement $U \to |A_{\uparrow\downarrow} (q,\omega)|$ in Eq.~(\ref{eq:resulttausd}). In Fig.~\ref{fig:fig2} we show the results for the spin-drag relaxation rate in two dimensions, as a function of temperature and for various values of the dimensionless parameter $\xi = \pi a/a_z$. Clearly, for sufficiently strong interactions, {\it i.e.}, sufficiently large $\xi$, this rate is enhanced upon approaching the ferromagnetic phase transition, as discussed in the introduction. For interactions that do not fulfill the Stoner criterion the spin-drag relaxation rate vanishes quadratically with temperature, as expected for a Fermi liquid. We note that, in two dimensions, there is a logarithmic correction to this quadratic temperature dependence \cite{damico2003} although this is hard to discern from the numerical results in Fig.~\ref{fig:fig2} and not the focus of this paper.

\begin{figure}[t]
\begin{center}
\includegraphics[width=0.6\linewidth]{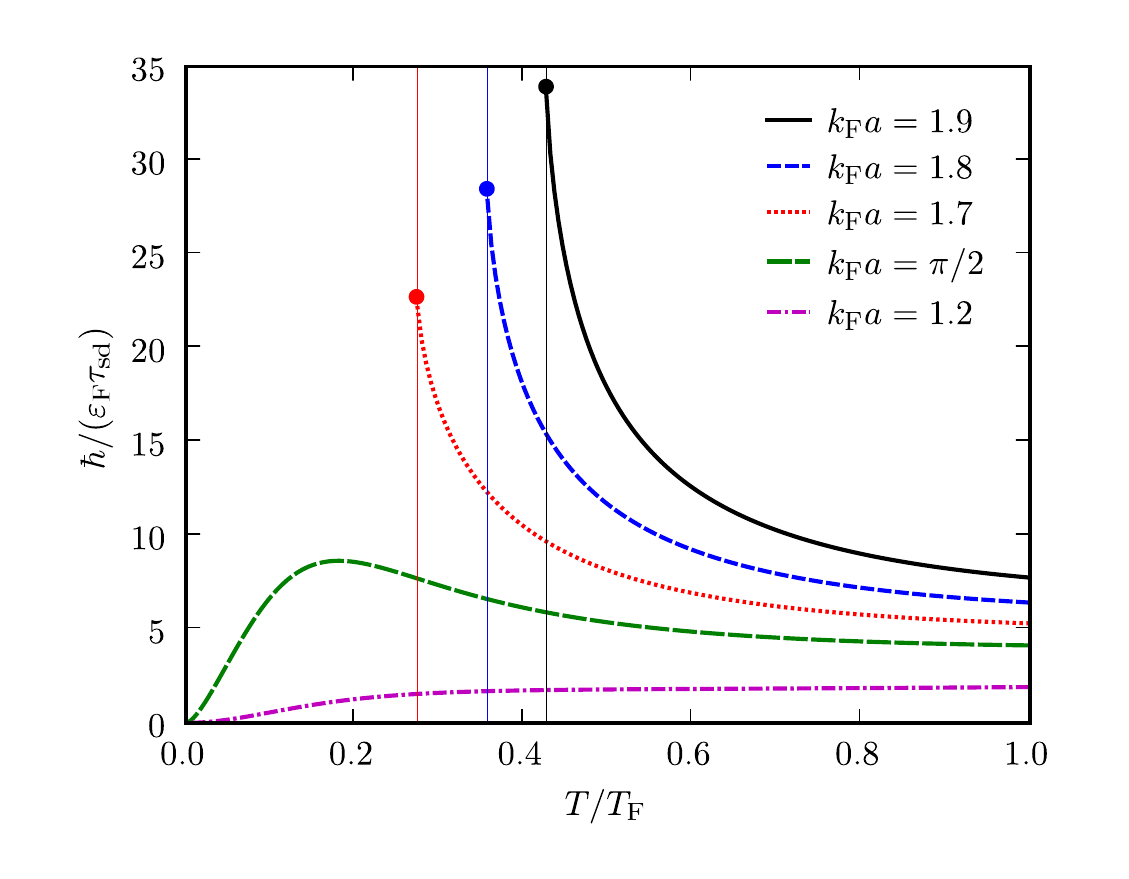}
\caption{Spin-drag relaxation rate of a three-dimensional Fermi gas as a function of temperature $T$, for various values of
the interaction parameter $k_{\rm F} a$. The vertical lines indicate the critical temperature for the ferromagnetic phase transition.\label{fig:fig3}}
\end{center}
\end{figure}

The three-dimensional results are obtained by replacing $U_{2D}$ with the three-dimensional two-body T-matrix $4\pi a\hbar^2/m$ \cite{duine2010}. The results for the spin-drag relaxation rate are shown in Fig.~\ref{fig:fig3} and are qualitatively similar to the two-dimensional case. For weak interactions, such that there is no ferromagnetic phase transition, the spin-drag relaxation rate vanishes quadratically with temperature.

\section{Discussion and conclusions} \label{sec:discussionandconclusions}
We have presented results for the spin-drag relaxation rate for a one-, two-, and three-dimensional two-component Fermi gas of ultracold atoms. In two and three dimensions, and for sufficiently strong interactions such a system may undergo a ferromagnetic phase transition. The spin-drag relaxation rate is strongly enhanced as this transition is approached from above, which could be observed experimentally as an increased damping of the spin dipole mode. This enhancement is determined by including all bubble-diagram contributions to the effective interaction between different spin components of the gas. This essentially treats the ferromagnetic phase transition within Stoner mean-field theory.  In three dimensions this is most likely qualitatively correct, although the transition occurs at strong coupling. In two dimensions the mean-field results for the critical temperature are an upper bound. This is because in this case, and in particular in the experimentally relevant case that an external field is present to trap two low-field seeking hyperfine species, the phase transition is of the Kosterlitz-Thouless type. It is known that the Stoner mean-field theory overestimates the Kosterlitz-Thouless transition temperature.

In one dimension the spin-drag relaxation rate vanishes linearly with temperature. In principle, we could also have included an effective interaction that included all bubble-like diagrams in the one-dimensional situation as well. This would result in an enhancement, at some temperature $T_{\rm SDW}$, of the spin-drag relaxation rate due to the divergence of the spin-density response function at zero frequency and $q=2k_{\rm F}$ that signals the onset of spin-density wave antiferromagnetism. This mean-field treatment, however, is not accurate in one dimension, not even qualitatively. Instead, it is known from bosonization theory that the linear dependence at small temperatures is the correct one. Since this behavior is reproduced by our expression in Eq.~(\ref{eq:resulttausd}) without including additional contributions to the  effective interactions, we do not include such fluctuation corrections in one dimension.

In future work we investigate the behaviour of the spin drag in the spontaneously spin-polarized phase, {\it i.e.}, for temperatures $T<T_{\rm c}$. Further studies will also investigate the role of critical fluctuations close to the critical temperature, and the situation of negative scattering length.

This work was supported by the Stichting voor
Fundamenteel Onderzoek der Materie (FOM), the Netherlands
Organization for Scientific Research (NWO), and by the European
Research Council (ERC). G.V. acknowledges support from NSF Grant
No. 0705460. M.P. acknowledges very useful conversations with
Rosario Fazio and Andrea Tomadin.

\end{document}